\documentclass[prl,twocolumn,showpacs,preprintnumbers,amsmath,amssymb]{revtex4}

\usepackage{graphicx}
\usepackage{dcolumn}
\usepackage{bm}

\begin{document}

\preprint{HgTe-low-field}

\title{Strained HgTe: a textbook 3D topological insulator}

\author{Cl\'ement Bouvier}
\affiliation{Institut N\'eel, C.N.R.S.- Universit\'e Joseph Fourier, BP 166,
38042 Grenoble Cedex 9, France}
\author{Xavier Baudry}
\affiliation{CEA, LETI, MINATEC Campus, DOPT, 17 rue des martyrs 38054 Grenoble Cedex 9, France}
\author{Philippe Ballet}
\affiliation{CEA, LETI, MINATEC Campus, DOPT, 17 rue des martyrs 38054 Grenoble Cedex 9, France}
\author{Tristan Meunier}
\affiliation{Institut N\'eel, C.N.R.S.- Universit\'e Joseph Fourier, BP 166,
38042 Grenoble Cedex 9, France}
\email[Corresponding author:]{tristan.meunier@grenoble.cnrs.fr}
\author{Roman Kramer}
\affiliation{Institut N\'eel, C.N.R.S.- Universit\'e Joseph Fourier, BP 166,
38042 Grenoble Cedex 9, France}
\author{Laurent P. L\'evy}
\affiliation{Institut N\'eel, C.N.R.S.- Universit\'e Joseph Fourier, BP 166,
38042 Grenoble Cedex 9, France}
\email[Corresponding author:]{Laurent.Levy@grenoble.cnrs.fr}
\homepage{http://www.neel.cnrs.fr/}

\date{\today}

\begin{abstract}
Topological insulators can be seen as band-insulators with a conducting surface.  The surface carriers are Dirac particles with an energy which increases linearly with momentum.  This confers extraordinary transport properties characteristic of Dirac matter, a class of materials which electronic properties are ``graphene-like''. We show how HgTe, a material known to exhibit 2D spin-Hall effect in thin quantum wells,\cite{Konig2007} can be turned into a textbook example of Dirac matter by opening a strain-gap by exploiting the lattice mismatch on CdTe-based substrates.  The evidence for Dirac matter found in transport shows up as a divergent Hall angle at low field when the chemical potential coincides with the Dirac point and from the sign of the quantum correction to the conductivity.  The material can be engineered at will and is clean (good mobility) and there is little bulk contributions to the conductivity inside the band-gap.
\end{abstract}

\pacs{85.25.Cp, 03.65.Vf, 74.50.+r, 74.78.Na} \keywords{Topological Insulators, Dirac matter, weak anti-localization}
 \maketitle

Graphene research\cite{Geim2007,DasSarma2011} has stimulated a considerable interest on Dirac matter, a novel class of materials where one or more bands have a Dirac-like dispersion in the vicinity of the Fermi level. Newly discovered materials show that Dirac matter can take a variety of forms. For example, angle-resolved photo-emission experiments have measured bands with a linear spectrum below the Fermi level in topological insulators surface states\cite{Hasan2010}, in organic conductors\cite{Katayama2006} and of course graphene stacks\cite{Rotenberg2006} with an odd number of layers.  Transport experiments give smoking-gun evidences for Dirac fermions in graphene. For organic-conductors and topological-insulators, other contributions make transport data more difficult to unravel.  This letter presents textbook evidences of Dirac fermions in a strained Mercury Telluride stack, using the low field magneto-transport behavior of a gated device.

\begin{figure}[h]
\includegraphics[width=0.9\columnwidth]{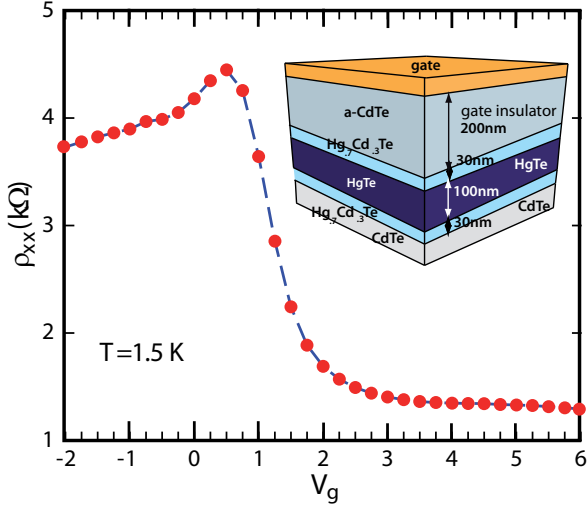}
\caption{Inset: the 100nm thick HgTe slab is sandwiched between two 30nm thick Hg$_{0.3}$Cd$_{0.7}$Te barriers.  A 200\:nm thick amorphous CdTe layer on top serves as a gate insulator.  Gate-voltage dependence of the resistivity.  $V_{g0}\approx 1V$ separates the electron and hole transport regimes: Above $V_{g0}$ transport is dominated by the surfaces while a coexistence with low-mobility heavy holes increases the resistivity below $V_{g0}$}\label{rhoxx-Vg}
\end{figure}

Mercury (HgTe) and Cadmium (CdTe) Telluride have the same zinc-blende structure.\cite{Bastard1984} However compared to CdTe, HgTe has a band inversion at the $\Gamma$ point: the $\Gamma_8$ bands which have a $P$ character lie 0.3 eV above the $\Gamma_6$ band (S character).  When growing epitaxially HgTe on top of CdTe, the HgTe lattice constant expands to match the CdTe lattice, which applies a lateral strain to HgTe.  As long as the HgTe thickness does not exceed a critical value above which dislocations appear, the strain is homogenous through the material.\cite{Bastard1984} A number of papers\cite{Fu2007} have pointed out that such strain opens a gap between the $\Gamma_8$ light and heavy hole bands, turning HgTe into a topological insulator.\cite{Matthews1974}  The value of this gap can be engineered to some extend in the epitaxial growth.  In this work, a symmetric Hg$_{0.3}$Cd$_{0.7}$Te/HgTe/Hg$_{0.3}$Cd$_{0.7}$Te stack (shown in the inset of Fig.~\ref{rhoxx-Vg}) was epitaxially grown on a CdTe $[211]$ face.  The direct gap between light and heavy holes bands is estimated to be of order 11\:meV while the indirect gap has been measured by thermal activation transport to be $\approx$6\:meV [supplementary material-a].  The $\Gamma_8$ and $\Gamma_6$ bands have a linear crossing\cite{Bastard1984} at the HgTe/Hg$_{0.3}$Cd$_{0.7}$Te interface (different symmetries) leading to 2D-``relativistic-like'' surface bands which band velocity $c$ plays the role of the speed of light.  These surface states are one of the manifestations of the non-trivial topology of HgTe\cite{Fu2007}.  The dispersion of surface state in a strained material have been analyzed qualitatively from the surface local density of states\cite{R-L-Chu2011} and in recent transport and magneto-optical experiments\cite{Brune2011,Hancock2011}. Nevertheless quantitative data is not yet available for our stack structure: it is however relatively simple to infer the surface band and the Dirac point positions by scanning the gate within the gap in transport measurements.

A 200$\:\mu$m long by 50$\:\mu$m wide six-probes gated device has been made by etching the stack and contacting {\em the top surface}.  The accessible range of gate voltages allows scanning part of the heavy-hole band (below V$_{g0}\approx$\:0.8\:V) and the gap region (V$_{g}$ between 1 and 6\:V).  An overview of the sample resistivity is shown in Fig.\ref{rhoxx-Vg}.  The resistivity is lowest ($\rightarrow\:1.2\:k\Omega$ at 6\:V) in the gap area where the surface conduction dominates the transport.  In the ``hole region'' (below 1\:V), there is a co-existence between surface and bulk hole conduction.   The mobility of heavy holes is low (large mass and short mean free path) leading to a higher resistivity ($\approx\:4\:k\Omega$) below 1\:V in spite of the higher carrier density.

\begin{figure}[b]
\centering
\includegraphics[width=0.8\columnwidth]{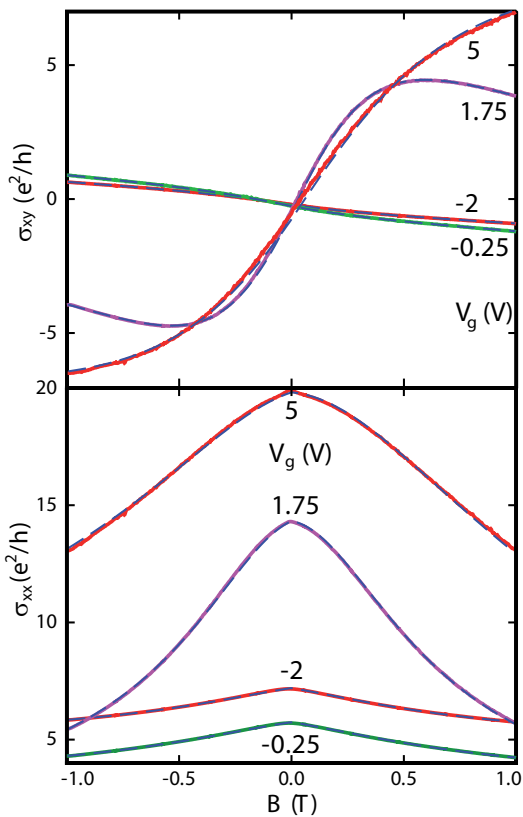}
\caption{Longitudinal  $\sigma_{xx}$ and Hall $\sigma_{xy}$ magnetoconductance expressed in units of $e^2/h$. The solid lines are obtained by inverting the measured resistivity tensor $(\rho_{xx},\rho_{xy})$ and rescaling in units of $e^2/h$.  The dashed lines are obtained by fitting the experimental curves to $\sigma^D_{xx}+\sigma_0$ and $\sigma^D_{xy}$.  The difference between the experimental curves and the fit are analyzed in terms of the quantum correction to the conductivity (weak-antilocalization) in Figure \ref{antilocalization}.}\label{magnetoconductance}
\end{figure}

The hole and electron character of the conduction can be read-off directly from the Hall conductivity plot shown in Fig.\ref{magnetoconductance}-top panel, where the slope at B=0 is negative at negative gate voltages and positive above V$_{g0}$.  Note also that the Hall slope gets larger as V$_{g0}$ is approached from below or from above.  These observations suggest an analysis of the magneto-transport using a two-fluid model: a surface contribution with a Dirac-like character and a bulk contribution.

The Lorentz force affects the motion of a relativistic charged particle with energy $\epsilon({\vec k})=\hbar c |{\vec k}|$ and momentum $\hbar {\vec k}$ in a similar fashion as for a massive charged carrier: the semiclassical equation of motion $\frac{d {\vec k}}{dt} =-\frac{ec}{\hbar} {\hat k} \times {\vec B}$ describes a cyclotron motion with angular frequency $\Omega_c = \frac{eBc^2}{\epsilon({\vec k})}$ which diverges at the Dirac point $\epsilon(k)\rightarrow 0$.  In the presence of the electric field $E{\hat x}$ induced by the bias voltage and of particle collisions (specified by the scattering time $\tau$), the average particle velocity $\langle {\vec v} \rangle = c \langle {\hat k} \rangle$ drifts at an angle $\Theta_{\rm Hall}$ with respect to the electric field
\begin{equation}
\tan \Theta_{\rm Hall} =\frac{\sigma_{xy}}{\sigma_{xx}} = \frac{\langle v_y \rangle}{\langle v_x \rangle} = \Omega_c \tau = \frac{2eD}{\mu} B = \frac{B}{B_*},
\end{equation}
where the particle energy has been replaced by the chemical potential $\mu$, $D= c^2 \tau/2$ is the Dirac particles diffusion coefficient, and the characteristic field $B_*=\mu/(2eD)$ is proportional to the chemical potential.  This Hall angle {\em diverges at the Dirac point} ($\mu\rightarrow 0$), a fundamental difference with ordinary particles where $\omega_c \tau = \frac{e\tau}{m} B$ does not depend on energy.

\begin{figure}[t]
\centering
\includegraphics[width=0.9\columnwidth]{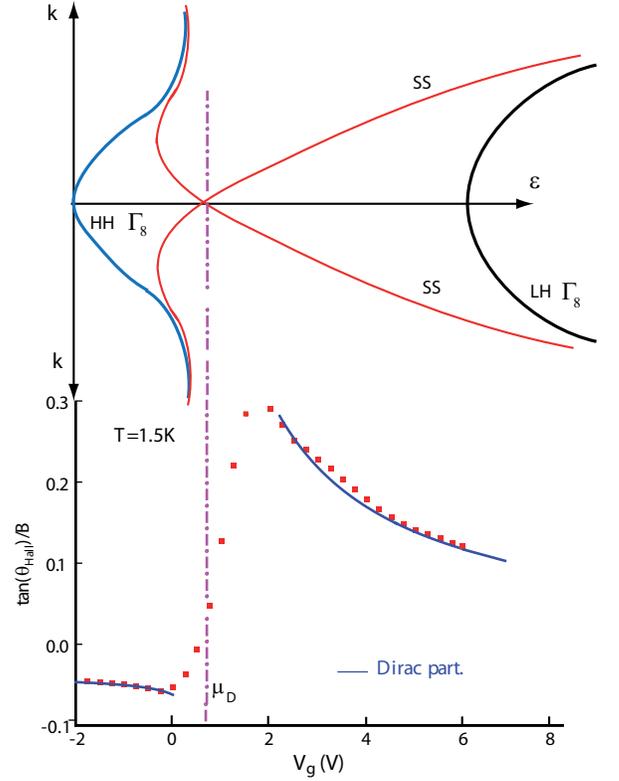}
\caption{Bottom: plot of the Hall angle slope as a function of the gate voltage.  An asymmetric cutt-off divergence is observed on each side of $V_{g0}$ consistent the expected magnetoconductance of Dirac particles. $V_{g0}$ is interpreted as the location of the Dirac point ($\mu_D$) in gate voltage.  The observed asymmetry is attributed to a difference in diffusion coefficients on the electron ($V_g>V_{g0}$) and the hole side where the bulk heavy hole band starts to be populated. Top: putting together the analysis of the magneto-transport and activation data, the inferred positions of the heavy (HH) and light (LH) hole bands with respect to the surface states (SS) connecting them.  The position of the Dirac point (V$_{g0}=0.8$V) can also be verified by extrapolating the observed $n=0$ quantum Hall state to zero magnetic field\cite{Bouvier2011}.) } \label{Hall-angle}
\end{figure}

\begin{figure}[t]
\centering
\includegraphics[width=0.9\columnwidth]{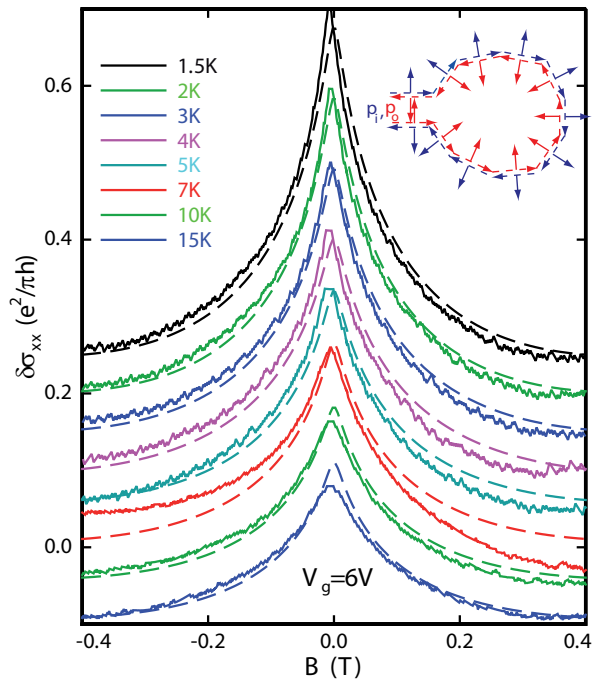}
\caption{The quantum correction to the conductivity are obtained by subtracting the two-fluid fit to the measured longitudinal conductivity. The difference are plotted as a function of magnetic field for different temperatures.  The curves are fitted to the expected digamma dependence as a function of field.  The characteristic field is $B_i=40\:$mT at T=1.5\:K and increases with increasing temperature.  Beyond $B_i$, the fitted curves (dotted lines) are dominated by the logarithmic tails expected in 2D.  }\label{antilocalization}
\end{figure}

The measured Hall angle slope ($\sigma_{xy}/(\sigma_{xx}B)$) is plotted on Fig.~\ref{Hall-angle}-bottom: a cut-off divergence is observed close to the gate voltage $V_{g0}$ with an asymmetry between the electron ($V_g>V_{g0}$) and the hole side.  This asymmetry comes from a partial population of bulk low-mobility heavy holes below $V_{g0}$ which also diffuse the Dirac particles because of their low mobilities.  By contrast, the electron side of the Dirac dispersion lies inside the strain-induced $\Gamma_{8}$ light-heavy holes gap. The observation of this divergent Hall angle, very similar to graphene\cite{Novosolov2004}, is a strong evidence for Dirac particles at the HgTe surfaces.
 A bulk hole conductivity $\sigma_0(V_g)$ appears below $V_{g0}$, which is nearly field independent at low (<1\:T) field ($\omega_c\tau_{HH}\ll 1$).  Hence, the experimental magnetoconductance is dominated by the surface Dirac particles, which reads [supplementary-material-b]
\begin{equation}
\sigma^D_{xx} = k_F \ell_e \frac{B_*^2}{B^2+B_*^2},~~~~\sigma^D_{xy} = k_F \ell_e \frac{B B_*}{B^2+B_*^2}, \label{magnetoDirac}
\end{equation}
when expressed in units of the quantum of conductance, $e^2/h$ ($\ell_e= c\tau$ is the mean free path, and $k_F=\mu/(\hbar c)$).  The longitudinal $\sigma_{xx}$ and Hall $\sigma_{xy}$ conductivities experimentally plotted in Fig.~\ref{magnetoconductance} are derived by inverting the measured resistivity tensor $(\rho_{xx},\rho_{xy})$ and expressing the data in units of $e^2/h$ [see a flash animation in the supplementary-material-c].  Fits to the two fluid model $\sigma^D_{xx}+\sigma_0$ and $\sigma^D_{xy}$, shown as dotted lines, are found to be quite accurate for all gate voltages.  The residual differences between the data and fits is attributed below to the quantum corrections to the conductivity (antilocalization shown in Fig.~\ref{antilocalization}).

A qualitative discussion of the data sheds some light on the origins of the parallel conduction $\sigma_0$.  At  V$_g\approx 1.75$V ($\approx 1$V above V$_{g0}$), the maximum of the Hall magnetoconductance occurs at B=0.6\:T.  In Eq.~\ref{magnetoDirac} this maximum is at $B_*=2eD/\mu(V_g)$ where the Hall conductance equals $k_F \ell_e/2$.  Since $B_*$ and $k_F \ell_e$ fully specify the Dirac magnetoconductance, the value of $\sigma_0$, the parallel conductance, is found to be of order one in units of $e^2/h$.  The value of $\sigma_0$ varies little above V$_{g0}$ but increases linearly in the hole region up-to $\approx 4.5 e^2/h {\\@}$ V$_g=-2\:$V.  In this region, $\sigma_0$ measures the gradual population of the bulk heavy hole subbands.  The non-zero V$_g$-independent value of $\sigma_0$ found in the gap region is more surprising.  In this experiment, the transport is measured by contacting {\em the top face} of the HgTe slab.  The sample has identical barriers on the top and bottom faces: the latter conduction is also expected to be dominated by 2D-Dirac carriers.  Based on high field data,\cite{Bouvier2011} we know that two faces are indeed connected through a \~25\:k$\Omega$ series-resistance\cite{comment1}, which explains the apparent gap conduction $\sigma_0$.

Strained HgTe appears to be an almost ideal topological insulator for transport studies: $\bullet$ the MBE growth yields easily gated devices with clean interface where mobilities
($\approx3\:10^4$cm$^ 2$/sec.) are already comparable to graphene; $\bullet$ the Dirac point lies in the gap, and the gap conduction is dominated by the surface Dirac carriers; $\bullet$ the bulk conduction is always very low, and in the coexistence region with the heavy hole band there is a natural mobility-selection of the surface carriers.

A resistance can be expressed as a probability of return to the origin of charge carriers.  In two dimensions, this probability strongly depends on closed-loops paths.   There are two ``time-reversed'' directions along which charge particles can travel along each closed loop.   For loop sizes smaller than the phase coherence length, the propagation amplitudes add coherently.  Depending on their relative phase, this can increase or decrease the probability of return to the origin.  For topological insulator surface states, the spin stays perpendicular to the momentum after a scattering event (spin-orbit interactions, see Fig.~\ref{antilocalization}-inset).  After the sequence of scattering on a closed loop, the spin has undergone a $2\pi$ rotation, which affects the accumulated phase (sign change):  For a given loop, the return probability is proportional to $\vert u_i({\vec p}, \uparrow)+\Theta u_i(-{\vec p}, \downarrow)\vert^2$, where $\Theta$ represents the time-reverse operation: since $\Theta u_i(-{\vec p},\downarrow)= - e^{2\pi \Phi/\Phi_0} u_i({\vec p},\uparrow)$,
the return probability is proportional to $\sin^2 2\pi\frac{\Phi}{\Phi_0}$.
The sign change comes from the spin rotation and the phase factor $e^{2\pi \Phi/\Phi_0}$ is the accumulated  Aharonov-Bohm phase along the loop ($\Phi$ is the flux through the loop and $\Phi_0=h/e$ the flux quantum).  The return probability is minimal at zero magnetic field, {\em i.e.} the quantum correction to the conductivity of the Dirac surface states are {\em negative}: this sign is opposite compared to ordinary conductors.  Such ``anti-localization'' quantum corrections to the conductivity  have been observed in graphene\cite{Wu2007} and other Dirac matter compounds\cite{Chen2010} and reveal the presence of a Dirac point\cite{McCann2006}.  These quantum corrections to the conductivity are obtained by subtracting the two-fluid fit to the measured longitudinal conductivity and plotted in Fig.~\ref{antilocalization}.  Their magnetic field dependence can be fitted to the known dependence\cite{McCann2006,Tkashov2011} [supplementary information-d].  At 1.5\:K, the magnitude of the weak-localization correction are $\approx 0.5$ times smaller than the expected magnitude $e^2/(2\pi h)$ for a perfect Dirac cone.  The characteristic field $B_i=B_{so}+B_\phi(T)$ has a temperature independent contribution $B_{so}\approx\:$40\:mT, which we attribute to an interfacial spin-orbit scattering length $\ell_{so}\approx 250\:$nm. For fields $B>B_i$, the expected long logarithmic tails dominate the field dependence.  In the hole region, the magnitude of quantum corrections drop by a factor of 2.5 and $B_i$ increase by the same factor as $\ell_{so}$ is reduced by the diffusion with heavy holes.

Strained HgTe appears to be a textbook realization of a 3D topological insulator.  Evidence for the Dirac point come from $\bullet$ the observed divergence of the Hall angle at the gate voltage $V_{g0}$, $\bullet$ the sign and field dependence of the quantum correction to the conductivity.  Within the strain-induced gap, the conduction is dominated by the 2D Dirac carriers.  In the coexistence region with the bulk heavy holes, the surface state conduction is reduced by the diffusion between surface and bulk carriers.  The processing and gating of this system is easy, making the fabrication of more complex structures (hybrids and spintronic devices) quite realistic.

This work was supported by the European GEOMDISS FET and ANR Xp-graphene contract. The authors are grateful to David Carpentier for enlightening discussions.


\begin{thebibliography}{99}
\bibitem[20]{comment1} The outer edge of the sample are chemically etched and are more disorded than the MBE grown HgTe/Hg$_{0.3}$Cd$_{0.7}$ interfaces.  Dislocations may also connect the top and bottom faces of the HgTe slab.
\end{thebibliography}
\end{document}